\documentclass[12pt,epsf,amssymb,qsymbols]{article}
\usepackage{tabularx}
\usepackage{array}
\usepackage{graphics}
\usepackage{graphicx}
\usepackage{epsfig}
\usepackage{amsmath}
\usepackage{amssymb}
\usepackage{ulem}
\makeatletter

\usepackage{verbatim}
\usepackage{citesort}

\newcommand{\msbar}{{\overline{\rm MS}}}

\newcommand{\bea}{\begin{eqnarray}}
\newcommand{\eea}{\end{eqnarray}}
\newcommand{\beq}{\begin{equation}}
\newcommand{\eeq}{\end{equation}}
\newcommand{\gev}{{\rm GeV}}
\newcommand{\mev}{{\rm MeV}}

\newcommand{\pdir}{p\kern -5.2pt\raise 0.2ex\hbox {/}}
\newcommand{\vdir}{v\kern -5.75pt\raise 0.15ex\hbox {/}}
\newcommand{\kdir}{k\kern -5.75pt\raise 0.15ex\hbox {/}}
\newcommand{\epsdir}{\epsilon\kern -5.0pt\raise 0.15ex\hbox {/}}
\newcommand{\bvdir}{\bar{v}\kern -5.75pt\raise 0.15ex\hbox {/}}
\newcommand{\Ddir}{D\kern -7.75pt\raise 0.20ex\hbox {/}}
\newcommand{\Adir}{A\kern -7.75pt\raise 0.20ex\hbox {/}}
\newcommand{\ldir}{l\kern -5.0pt\raise 0.2ex\hbox{/}}
\newcommand{\varepsdir}{\varepsilon\kern -5.5pt\raise 0.15ex\hbox{/}}

\makeatother

\begin{document}
\thispagestyle{empty} 
\begin{flushright}
\begin{tabular}{l}
{\tt RM3-TH/03-2} \\
{\tt Roma, 1351/03 }\\
{\tt SHEP-02-34}\\
\end{tabular}
\end{flushright}
\begin{center}
\vskip 1.2cm\par
{\par\centering \Large \bf Coupling of the light vector meson to the vector}\\
{\par\centering \Large \bf and to the tensor current}\\
\vskip 0.9cm\par
{\par\centering \large  
\sc D.~Becirevic$\,^{a,b}$, V.~Lubicz$\,^c$, F.~Mescia$\,^d$ and C.~Tarantino$\,^c$}
{\par\centering \vskip 0.5 cm\par}
{\sl
$^a$ Laboratoire de Physique Th\'eorique (B\^at 210)~\footnote{Unit\'e mixte de
Recherche du CNRS - UMR 8627.}, Universit\'e de Paris Sud, \\ 
Centre d'Orsay, 91405 Orsay-Cedex, France. \\
\vspace{.25cm}
$^b$ Dip. di Fisica, Universit\`a di Roma ``La Sapienza",\\
Piazzale Aldo Moro 2, I-00185 Rome, Italy. \\
\vspace{.25cm}
$^c$ Dip. di Fisica, Univ. di Roma Tre and INFN,
Sezione di Roma III, \\
Via della Vasca Navale 84, I-00146 Rome, Italy.\\
\vspace{.25cm}
$^d$ Department of Physics and Astronomy, University of Southampton,\\
Southampton SO171BJ, United Kingdom.}\\
%\vskip1.cm

\end{center}

%\vskip 0.25cm
\begin{abstract}
We present results for the coupling of the 
light vector mesons to the tensor current, relative to the standard 
vector meson decay constants.  
From an ${\cal O}(a)$-improved lattice study, performed at three values of the 
lattice spacing in the quenched approximation, our final values (in the continuum limit), 
in the $\msbar$ scheme at $\mu =2$~GeV, are: 
$f_\rho^T/f_\rho =0.72(2)(^{+2}_{-0})$, $f_{K^\ast}^T/f_{K^\ast} = 0.74(2)$, 
$f_{\phi}^T/f_{\phi} =0.76(1)$. 
\end{abstract}
\vskip 2.2cm
\setcounter{page}{1}
\setcounter{footnote}{0}
\setcounter{equation}{0}
%%%%%%%%%%%%%%%%%%%%%%%%%%%%%%%%%%%%%%%%

\renewcommand{\thefootnote}{\arabic{footnote}}
\vspace*{-1.5cm}
{
%\newpage
\setcounter{footnote}{0}
%%%%%%%%%%%  Section 1

\section{Introduction}
\setcounter{equation}{0}
\subsection{Motivation for this computation}
The most direct way to extract one of the least known CKM parameters, $\vert V_{ub}\vert $, is from 
studies of the corresponding leptonic and semileptonic decays of $B$-mesons. 
Although the leptonic decays are still beyond the reach of the present experiments, 
CLEO, Belle and BaBar are providing rather accurate measurements of the branching ratios for 
the semileptonic $B\to \rho (\pi) \ell \nu$ modes~\cite{experiment}. On the theoretical 
side, lattice QCD and light cone QCD sum rules (LCSR) are expected
to provide model independent information about   
shapes and absolute values of the relevant form factors.  
That necessitates good control over the low energy (non-perturbative) QCD dynamics. 
The complicating feature of $heavy\to light$ decays is that the $q^2$-region accessed by 
these decays is large, e.g. for $B\to \rho \ell \nu$, with $\ell = e, \mu$, it is $0 \leq q^2 
\leq (m_B-m_\rho)^2\equiv 20.3 \ \gev^2$. Lattice QCD results are available 
for large momentum transfers, $q^2 \geq 10 \ \gev^2$ (see refs.~\cite{lattice-btorho}), 
while LCSR results are expected to be reliable in the region of low $q^2$'s~\cite{CZ,bb1,bb2}.
In order to reduce the uncertainties in the form factors obtained in the latter approach, 
one needs better control over the non-perturbative parameters which are explicitly 
present in the LCSR, such as moments of the light cone wave functions, various hadronic couplings etc.  
A step in this direction is made in this letter, where we report results for the 
ratio of the coupling constants of the light vector mesons to the tensor ($f_V^T$)
and to the vector current ($f_V$), which is an important ingredient  
in the LCSR expressions for the $B\to \rho$ and $B\to K^\ast$ semileptonic form factors. 
While the constant $f_V$ can in principle be extracted from experiments on $e^+e^- \to V^0$ and 
$\tau^- \to V^- \nu_\tau$, the coupling $f_V^T$ can be estimated only theoretically. 
To do so, we will use the lattice QCD. The first computation of this coupling has been attempted 
in ref.~\cite{APE}. The QCDSF collaboration has also presented their preliminary results in 
ref.~\cite{QCDSF}. 
Here we will restrict our attention to the ratio of the couplings $f_V^T/f_V$, 
use the data generated at three values of the lattice spacing and extrapolate to the continuum limit.
Before we enter into details of the lattice computations, let us make a brief assessment of the 
importance of having an accurate determination of  $f_V^T/f_V$.

\subsection{How important are the couplings $f_V$ and $f_V^T$?}

\begin{figure}
\vspace*{-0.8cm}
\begin{center}
\begin{tabular}{@{\hspace{-0.25cm}}c}
\epsfxsize9.2cm\epsffile{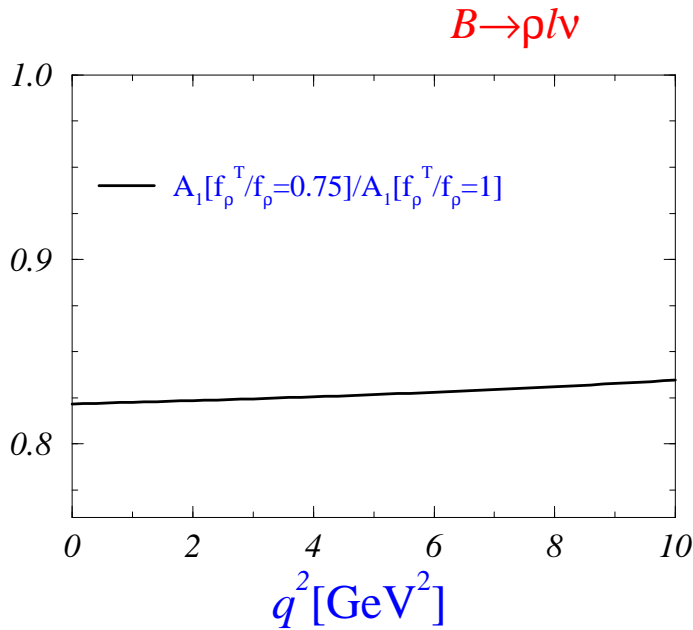}    \\
\end{tabular}
\vspace*{-.5cm}
%%%%%%%%%%%%%%%%%%%%%%%%%%%%%%%%%%%%%%%%%%%%%%%%%%%%%%%%%%%%%%%%%%
\caption{\label{fig1}{\footnotesize 
The effect of replacing $f_\rho^T/f_\rho =1 \to 0.75$ in eq.~\eqref{lcsr} results in about $15-20$\%
suppression of the form factor $A_1^{B\to \rho}(q^2)$. The ratio is evaluated by using the leading twist distribution 
amplitudes from ref.~\cite{bb2}.
} }
%%%%%%%%%%%%%%%%%%%%%%%%%%%%%%%%%%%%%%%%%%%%%%%%%%%%%%%%%%%%%%%%%%
\end{center}
\end{figure}
%%%%%%%%%%%%%%%%%%%%%%%%%%%%%%%%%%%%%%%%%%%%%%%%%%%%%%%%%%%

To exemplify the importance of having a good handle on $f_V^T/f_V$, we give the 
LCSR expression for the form factor which dominates the $B\to \rho \ell\nu$ decay rate, namely~\cite{bb1}
\bea \label{lcsr}
&& A_1(q^2) ={ m_b f_\rho {\rm e}^{(m_B^2-m_b^2)/M^2 } \over m_B^2 f_B (m_B+m_\rho) }
\int_{u_0}^1 {d u \over u} {\rm e}^{ (1-1/u)(q^2 - m_b^2 - u m_\rho^2)/M^2} \times \nonumber \\
&& 
\left\{  {m_b^2 + u^2 m_\rho^2 -q^2 \over 2 u} {f_\rho^T \over f_\rho} \phi_T(u) + 
{ m_b m_\rho\over 2} \left[\int_0^u dv{\phi_\parallel (v)\over 1-v} + \int_u^1 dv{\phi_\parallel (v)\over v}
\right]\right\}\,,
\eea
where $M^2$ is the so-called Borel parameter. The lower limit in the integral, $u_0$, contains  information 
about the energy of a threshold, $s_0$, above which quark-hadron duality is 
assumed. Using the input parameters of ref.~\cite{bb1}, at $q^2= 0$ one has $u_0\simeq 0.65$, whereas 
for   $q^2= 10\ \gev^2$ it is $u_0\simeq 0.5$. The objects under integrals are the light cone 
distribution amplitudes, which are the functions of  $u$, the  fraction of the longitudinal momentum 
of the meson carried by one of the meson's valence quarks. Indices $T$ and $\parallel$ denote the 
polarization states of the $\rho$-meson. In the asymptotic limit, $\mu\to \infty$, these functions are
$\phi^{\rm asymp.}_{T,\parallel} (u) = 6 u (1-u)$~\cite{CZ}. QCD corrections to 
$\phi^{\rm asymp.}_{T,\parallel} (u)$ are conveniently accounted for by  
expanding these functions in the basis of Gegenbauer polynomials and then by computing the first 
few moments. In order to make the above sum rule~\eqref{lcsr} as simple as possible 
we did not include radiative nor higher twist corrections. They can be found 
in ref.~\cite{bb2}.

The first term in eq.~\eqref{lcsr}, which is numerically dominant, is proportional to $f_\rho^T/f_\rho$.  
To show the importance of that ratio we first set it to one, 
as has been done in~\cite{CZ}, and then to $f_\rho^T/f_\rho \approx 0.75$, as typically found in 
 recent QCD sum rule analyzes~\cite{bb,bm}. We plug those values in eq.~\eqref{lcsr}, 
 make the ratio of the two cases and plot it in 
fig.~\ref{fig1}. From that plot and by varying the parameters as in ref.~\cite{bb1}, 
we see that the form factor $A_1(q^2)$ gets shifted by  $15$--$20$\%
. Therefore a reliable estimate of $f_\rho^T/f_\rho$, as to reduce the overall uncertainty on 
$A_1(q^2)$, is important.

\section{Definitions. Lattice details. Direct results}

\subsection{Vector meson couplings}
For a generic vector meson $V$, with valence quark content   
$(\bar q^\prime q)$, the couplings $f_V$ and $f^T_V(\mu)$ are 
defined through the following matrix elements
\bea \label{def1}
&&\langle 0\vert \bar q (0) \gamma^\mu q^\prime (0)
\vert V(p,\lambda) \rangle = f_V m_V e^\mu_\lambda \;, \nonumber \\
&& \\
&&\langle 0\vert \biggl( \bar q (0) \sigma^{\mu \nu} q^\prime(0)\biggr) (\mu)
\vert V(p,\lambda) \rangle = i f_V^T(\mu) \left( 
e^\mu_\lambda p^\nu - e^\nu_\lambda p^\mu \right) \;, \nonumber 
\eea
where $p$ and $e^\mu_\lambda$ are the momentum and the polarization vectors. 
These definitions refer to Minkowski space where we take 
$\sigma^{\mu \nu}=(i/2)[\gamma^\mu, \gamma^\nu]$. Notice that since the 
anomalous dimension of the tensor current is different from zero, the 
coupling $f^T_V(\mu)$ depends on the scale at which the corresponding 
current is renormalized.

\subsection{Relevant correlation functions}

To estimate the values of the coupling $f^T_V(\mu)$ for various light vector mesons $V$, 
we will compute its ratio with $f_V$. This is beneficial because several systematic 
uncertainties, such as the dependence on the lattice spacing, are likely to cancel to a large extent. 
To compute  $f^T_V(\mu)/f_V$ it is sufficient to consider the vector meson at rest ($\vec p = 0$) 
and the following ratio of correlation functions in Euclidean space (``$E$" in the following)
\bea
\label{r1}
R(t;\mu) = {C_{TV}(t)\over C_{VV}(t)} =
{\langle {\displaystyle \sum_{\vec x} }  \hat T_{0i}(x;\mu) V^\dagger_{i}(0) \rangle_E
\over
\langle {\displaystyle \sum_{\vec x} }  \hat V_{i}(x) V^\dagger_{i}(0)\rangle_E
}\;,
\eea
where on the r.h.s. we chose the second operator to be the vector current. Any other local 
operator with the quantum numbers $J^{PC}=1^{--}$ would also be a good choice. 
The hat symbol indicates that the corresponding 
current has been improved and renormalized, which we discuss in the next subsection. 
By inserting a complete set of states and going towards large time separations between the source operators, 
the lowest meson state dominates and we
have
\bea
\label{r2}
\biggl.R(t;\mu) \biggr|_{t\gg 0} = { {\displaystyle \sum_{\lambda} } e^\ast_{\lambda,i}\langle 0\vert \hat T_{0i}(\mu) \vert V(0,\lambda)\rangle_E  \over
{\displaystyle \sum_{\lambda} } e^\ast_{\lambda,i} \langle 0\vert \hat V_{i} \vert V(0,\lambda)\rangle_E  } \;.
\eea
In eq.~(\ref{r2}), we have Euclidean operators, whereas the couplings, as defined 
in eq.~(\ref{def1}), refer to  Minkowski space (``$M$"). Using the 
fact that 
\bea
T_{0 i}^E =(i\bar q(x) \sigma_{0i} q(x))^E\to (\bar q (x)\sigma^{0i} q(x))^M \;,\quad  
V_{i}^E =(\bar q(x) \gamma_{i} q(x))^E \to  -i(\bar q(x) \gamma^{i} q(x))^M \;,
\eea 
it becomes clear that the plateau in the ratio~\eqref{r1}, (i.e. eq.~\eqref{r2}) gives $R(\mu) = f_V^T(\mu)/f_V$.

\subsection{Improvement and renormalization}

Our lattice study is made by using non-perturbatively ${\cal O} (a)$ improved Wilson fermions, 
where $a$ stands for the lattice spacing. The improvement of the  
operators considered in this letter, $\hat T_{0i}(\mu)$ and $\hat V_i$, is made through~\cite{luscher1}
\bea \label{tt}
&&\hat T_{0 i}^E(\mu) = 
Z_T^{(0)}(\mu a) \bigl( 1 + b_T \ am_q \bigr) \biggl[ 
T_{0 i}^E(a) + c_T(a) \partial_0 V_i(a) \biggr]\;, \\
&& \hat V_i^E  = 
Z_V^{(0)}(a) \bigl(1 + b_V \ am_q  \bigr) \biggl[ 
V_i^E(a) - c_V(a) \partial_0 T_{0 i}^E(a) 
\biggr]\;. 
\eea 
$c_{T,V}(a)$ ensure that bare currents computed on the lattice are free of  
${\cal O} (a)$ lattice artifacts;  
the constants $Z_V^{(0)}(a)$ and $Z_T^{(0)}(\mu a)$ provide the matching to the continuum 
local operators and also their renormalization in the chiral limit, whereas the quark mass 
dependent artifacts of ${\cal O}(am_q)$ are subtracted by setting the coefficients $b_{V}(a)$ and  $b_{T}(a)$
to their values determined non-perturbatively and by using boosted perturbation theory, respectively. 
The numerical values of all of the above constants are listed 
in table~\ref{tab:1}, where we also give the main features 
of our lattices~\footnote{Please  
note that our lattice data refer to the same runs as those used in refs.~\cite{qmass,rcs}, 
where additional information can be found.}.
\begin{table}[h!]
\centering 
\begin{tabular}{|c|ccc|}  \hline \hline
{\phantom{\huge{l}}}\raisebox{-.2cm}{\phantom{\Huge{j}}}
$ \beta = 6/g_0^2$&  6.0 &  6.2 & 6.4     \\ 
{\phantom{\huge{l}}}\raisebox{-.2cm}{\phantom{\Huge{j}}}
$ c_{SW}$~\cite{csw} &   1.769 &  1.614 & 1.526  \\ 
{\phantom{\huge{l}}}\raisebox{-.2cm}{\phantom{\Huge{j}}}
$ L^3 \times T $&  $16^3 \times 52$ & $24^3 \times 64$  & $32^3 \times 70$ \\ 
{\phantom{\huge{l}}}\raisebox{-.2cm}{\phantom{\Huge{j}}}
$ \#\ {\rm conf.}$& 500 &  200 & 150  \\ \hline 
{\phantom{\huge{l}}}\raisebox{-.2cm}{\phantom{\Huge{j}}}
$\kappa_1$& 0.1335 & 0.1339 & 0.1347   \\ 
{\phantom{\huge{l}}}\raisebox{-.2cm}{\phantom{\Huge{j}}}
$\kappa_2$& 0.1338 & 0.1344 &  0.1349  \\ 
{\phantom{\huge{l}}}\raisebox{-.2cm}{\phantom{\Huge{j}}}
$\kappa_3$& 0.1340 & 0.1349 &  0.1351  \\ 
{\phantom{\huge{l}}}\raisebox{-.2cm}{\phantom{\Huge{j}}}
$\kappa_4$& 0.1342 & 0.1352 &  0.1353  \\ \hline 
{\phantom{\huge{l}}}\raisebox{-.2cm}{\phantom{\Huge{j}}}
$a/r_0$~\cite{necco} &   0.1863 & 0.1354 & 0.1027     \\ \hline 
{\phantom{\huge{l}}}\raisebox{-.2cm}{\phantom{\Huge{j}}}
$c_T$~\cite{LANL}& 0.07 & 0.06   & 0.05  \\ 
{\phantom{\huge{l}}}\raisebox{-.2cm}{\phantom{\Huge{j}}}
$c_V$~\cite{LANL}& -0.11 & -0.09   & -0.08  \\ 
{\phantom{\huge{l}}}\raisebox{-.2cm}{\phantom{\Huge{j}}}
$b_V$~\cite{alphabM} & 1.47& 1.41 & 1.36   \\
{\phantom{\huge{l}}}\raisebox{-.2cm}{\phantom{\Huge{j}}}
$b_T$~\cite{capitani} & 1.23&  1.22 & 1.21   \\
{\phantom{\huge{l}}}\raisebox{-.2cm}{\phantom{\Huge{j}}}
$Z_{V}^{(0)}(a)$~\cite{rcs} &  0.766(2) & 0.775(2) & 0.795(3) \\
{\phantom{\huge{l}}}\raisebox{-.2cm}{\phantom{\Huge{j}}}
$Z_{T}^{(0)}(\mu a= 1)$~\cite{rcs} &  0.833(2) & 0.847(3) & 0.867(6) \\
 \hline \hline
\end{tabular}
%%%%%%%%%%%%%%
{\caption{\footnotesize  \label{tab:1} Summary of our lattice details. 
We also give the values of the improvement coefficients (with the corresponding references), 
and the values of the lattice spacing relative to the parameter $r_0\approx 0.5$~fm. }}
\end{table}

\subsection{Direct lattice results}

Information about our lattices is provided in table~\ref{tab:1}. 
Our data is obtained at three lattice spacings. The improvement ensures 
that the effects linear in lattice spacing are not present and 
therefore the extrapolation to the continuum limit ($a\to 0$) is expected to
be smoother. In table~\ref{tab:2} we collect results directly obtained 
for our lattices.
\begin{table}[h!]
\begin{center} 
\begin{tabular}{c|c|c|c|c|c|}
\cline{3-6}
\multicolumn{2}{l|}{}&{\phantom{\Huge{l}}}\raisebox{-.1cm}{\phantom{\Huge{j}}}
\underline{$\kappa_1$}& \underline{$\kappa_2$}& \underline{$\kappa_3$}& \underline{$ \kappa_4 $}  \\
 \cline{2-6} 
{\phantom{\Huge{l}}}\raisebox{-.1cm}{\phantom{\Huge{j}}}
\uwave{$\beta=6.0$} & $m_P/m_V$   & 0.743(6) & 0.709(8) & 0.682(10) &  0.650(12) \\  
{\phantom{\Huge{l}}}\raisebox{-.1cm}{\phantom{\Huge{j}}}
            & $f_V^T/f_V$ & 0.739(5) & 0.733(7) & 0.730(10) & 0.728(14) \\ \cline{2-6} 
{\phantom{\Huge{l}}}\raisebox{-.1cm}{\phantom{\Huge{j}}}
\uwave{$\beta=6.2$} & $m_P/m_V$   & 0.807(4) & 0.751(7) & 0.661(11) &  0.575(16) \\  
{\phantom{\Huge{l}}}\raisebox{-.1cm}{\phantom{\Huge{j}}}
            & $f_V^T/f_V$ & 0.761(4) & 0.744(5) & 0.731(9) & 0.736(15) \\ \cline{2-6} 
{\phantom{\Huge{l}}}\raisebox{-.1cm}{\phantom{\Huge{j}}}
\uwave{$\beta=6.4$} & $m_P/m_V$   & 0.750(8) & 0.707(9) & 0.648(11) &  0.563(15) \\  
{\phantom{\Huge{l}}}\raisebox{-.1cm}{\phantom{\Huge{j}}}
            & $f_V^T/f_V$ & 0.747(4) & 0.737(5) & 0.728(7) & 0.723(15) \\ \cline{2-6} 
\end{tabular}
%\vspace*{.8cm}
\caption{\label{tab:2}
\footnotesize  Ratio of the masses of the pseudoscalar and vector mesons consisting of 
two degenerate quarks of mass corresponding to $\kappa_i$ specified in table~\ref{tab:1}.
We also give the values of the $f_V^T/f_V$ ratio for each $\kappa_i$, where the renormalization scale 
is set to $\mu=(1/a)_\beta$.}
\end{center}
\vspace*{-.3cm}
\end{table}

In forming the ratio~\eqref{r1} we used time reversal to symmetrise the correlation functions as
\bea
C_{TV}(t) &\to& {1\over 2} \left[ C_{TV}(t) - C_{TV}(T-t) \right] \;,\cr
C_{VV}(t) &\to & {1\over 2}\left[ C_{VV}(t) + C_{VV}(T-t) \right] \;,
\eea
where $t=0,1,\dots T/2$. On the plateaus, the symmetrised correlator produces the usual 
$\tanh$-factor, and thus the final fit form is
\bea \label{ratio-def}
R(t;1/a) = {f_V^T \over f_V} \tanh\left[ m_V \left( T/2 - t \right)  \right]\;.
\eea
In improving the currents (see eq.~\eqref{tt}) we use the symmetric discrete derivative i.e. 
$\partial_0 g(t) = [g(t+1) -g(t-1)]/2$, where $g(t)$ is the generic Green function 
computed on the lattice.

\begin{figure}
\vspace*{-0.8cm}
\begin{center}
\begin{tabular}{@{\hspace{-0.25cm}}c}
\epsfxsize10.2cm\epsffile{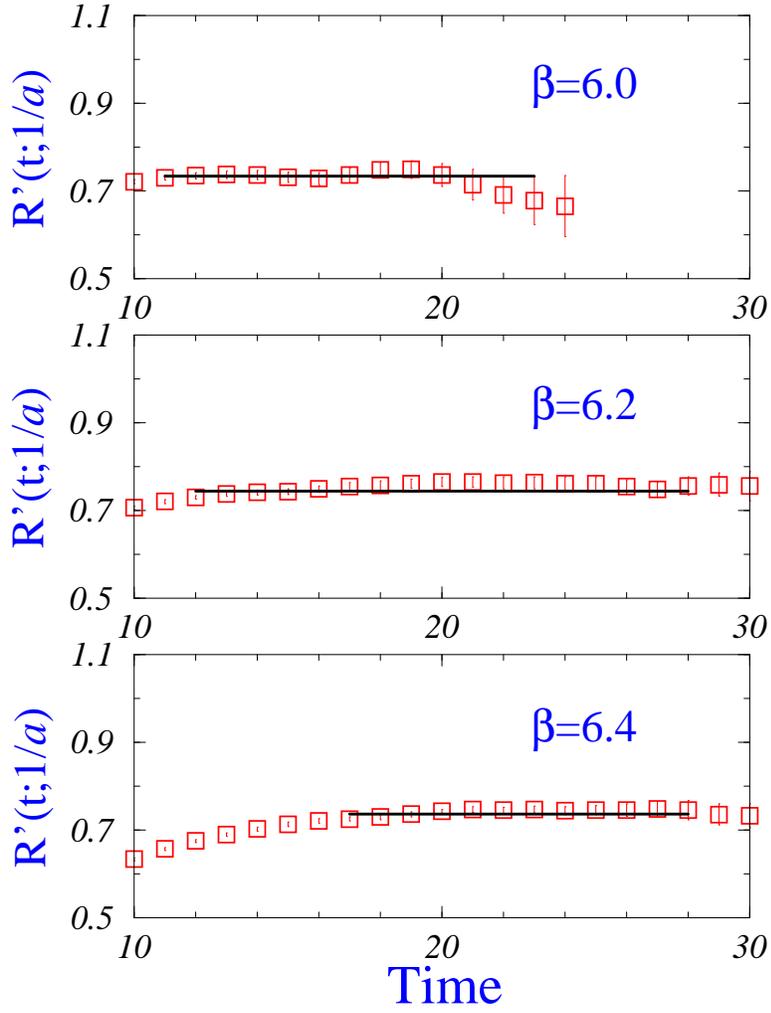}    \\
\end{tabular}
\vspace*{-.1cm}
%%%%%%%%%%%%%%%%%%%%%%%%%%%%%%%%%%%%%%%%%%%%%%%%%%%%%%%%%%%%%%%%%%
\caption{\label{fig2}{\footnotesize 
Ratio $R'(t;1/a)=R(t;1/a)/\tanh\left[ m_V \left( T/2 - t \right)\right]$ 
(see eqs.~(\ref{r1},\ref{ratio-def})) which on the plateau leads to $f_V^T/f_V$. 
Fit to a constant is also shown. We show the signals for all three values 
of the lattice spacing, and we choose $\kappa=\kappa_2$ specified in 
table~\ref{tab:1} for each $\beta$.
 } }
%%%%%%%%%%%%%%%%%%%%%%%%%%%%%%%%%%%%%%%%%%%%%%%%%%%%%%%%%%%%%%%%%%
\end{center}
\end{figure}
%%%%%%%%%%%%%%%%%%%%%%%%%%%%%%%%%%%%%%%%%%%%%%%%%%%%%%%%%%%

Finally, the fit is made using the same intervals with which the vector meson masses have been extracted
(see ref.~\cite{qmass}), namely
\bea
\beta=6.0 &\to& t\in[11,23]\;,\cr
&&\cr
\beta=6.2 &\to&  t\in[12,28]\;,\cr
&&\cr
\beta=6.4 &\to&  t\in[17,28]\;.
\eea 
The results are listed in table~\ref{tab:2}, and an illustration of the ratios 
$R(t;1/a)$ is provided in fig.~\ref{fig2}.

\section{Physical results}

\subsection{Running}
We first need to run all the ratios from the scale $\mu = 1/a$ to the conventional 
$\mu = 2$~GeV. In our computation we used the constant $Z_T(\mu)$ determined 
non-perturbatively in the (Landau)RI/MOM renormalization scheme~\cite{rcs} which, 
up to NLO in perturbation theory, is the same scheme as $\msbar$(NDR). The scale dependence of 
the tensor bilinear, and thus of $Z_T(\mu)$, is obtained by solving 
the renormalization group equation at NLO:
\bea \label{rge}
{d \log \langle T_{\mu \nu}(\mu)\rangle \over d\log \mu } = - \gamma_T\left(\alpha_s(\mu)\right)\;,
\eea
where the anomalous dimension at NLO in perturbation theory reads~\cite{gracey}
\bea
\gamma_T\left(\alpha_s(\mu)\right) =
\gamma_T^{(0)} {\alpha_s(\mu)\over 4 \pi } +  \gamma_T^{(1)}\left( {\alpha_s(\mu)\over 4 \pi }\right)^2
= {8\over 3} {\alpha_s(\mu)\over 4 \pi } + 
{4\over 27}\left(543 - 26 n_F\right) \left( {\alpha_s(\mu)\over 4 \pi }\right)^2
\;.
\eea
At the same order, the running coupling is governed by
\bea
{d \alpha_s(\mu) \over d\log \mu } = - 2 \beta_0 \alpha_s(\mu) - {\beta_1\over 2 \pi} \alpha_s^2(\mu) \;,
\eea
with the beta function coefficients being $\beta_0 = 11 -2 n_F/3$, and 
$\beta_1 = 102 -38 n_F/3$. 
Since $f_V^T(\mu)$ inherits the scale dependence from the tensor operator, it is 
completely equivalent to solve eq.~\eqref{rge} in terms of $f_V^T(\mu)$ instead of 
the matrix element $\langle T_{\mu \nu}(\mu)\rangle$. We finally have
\bea \label{evol}
f_V^T(\mu) &=& \nonumber f_V^T(\mu_0) W[\mu,\mu_0] \\ 
&\equiv&f_V^T(\mu_0) \left( {\alpha_s(\mu)\over
\alpha_s(\mu_0)}\right)^{4/(33 - 2 n_F)} \times \left[ 1\  +\ J_T\  {\alpha_s(\mu)-\alpha_s(\mu_0) \over 4\pi } 
\right]\;,  \\ 
&&\cr
{\rm where} && \quad J_T\ =\  { \gamma_T^{(1)} \beta_0 - \gamma_T^{(0)} \beta_1 \over 2 \beta_0^2 } \ = \ {2\over 9}\
{12411-1260 n_F+52 n_F^2\over (33 - 2 n_F)^2}\ \;.\nonumber
\eea
Since we work in the quenched approximation, we set $n_F=0$, and we run 
to $\mu =2$~GeV by using $\Lambda_\msbar^{n_F=0}=0.250(25)$~GeV.
The values of the lattice spacing are computed 
by using the conservative estimate $a^{-1}_{\beta=6.0} = 2.0(1)$~GeV. 
The other $a^{-1}_{\beta}$ are related to $a^{-1}_{\beta=6.0}$ by using the ratios of $a/r_0$~\cite{necco}.  
The net effect is that
\bea
W[2\ \gev,a^{-1}] = \Bigl. 1.000(2)(0) \Bigr|_{\beta=6.0} , \Bigl.1.010(2)(1) \Bigr|_{\beta=6.2},
\Bigl.1.019(2)(1) \Bigr|_{\beta=6.4},
\eea
where the first error comes from the $5\%
$ variation of $a^{-1}_{\beta=6.0}$, and the second one from $10\%
$ uncertainty on $\Lambda_\msbar^{n_F=0}$. 
As we can see those uncertainties are a few per mil and are thus negligible.

\subsection{ $f_\rho^T(2 \ \gev)/f_\rho$, $f_{K^\ast}^T(2 \ \gev)/f_{K^\ast}$ and $f_{\phi}^T(2 \ \gev)/f_{\phi}$ }

To get the physically relevant results (corresponding to $V=\rho$, $K^\ast$, $\phi$), we fit 
our results to the form 
\bea\label{extr}
{ f_V^T(2 \ \gev)\over f_V} = \alpha_0   + \alpha_1 ( a m_P )^2  + \alpha_2 (a m_P)^4\;,
\eea
where $(a m_P)$ stands for the pseudoscalar meson mass that is computed by using the same 
quarks (non-degenerate in mass) as the ones used to compute the ratio on the l.h.s. (see  
table~\eqref{tab:2}). As our central values, we chose to quote the results obtained through 
the linear fit ($\alpha_2 =0$), whereas the quadratic fit is used to assess  
systematic uncertainties. The physical results, corresponding to $V=\rho$, $K^\ast$, $\phi$, 
are obtained by choosing $(a m_P)^2 = (a m_\pi)^2$, $(a m_K)^2$ , 
$[2(a m_K)^2-(a m_\pi)^2]$, respectively. The values of $(a m_\pi)^2$ and $(a m_K)^2$ are obtained from the 
fit of our data to
\bea
(a m_V) = \delta_0  + \delta_1 ( a m_P )^2\;,
\eea
and the intersection of this form with the curve $(a m_V) = r \sqrt{( a m_P )^2}$, where 
$r = 5.6, 1.8$, i.e. $(m_\rho/m_\pi)_{\rm phys.}$, and  $(m_{K^\ast}/m_{K})_{\rm phys.}$, 
respectively. This extrapolation-interpolation to physical mesons is 
shown in fig.~\ref{fig:3} and the results are presented in table~\ref{tab:3}. 
\begin{figure}
\vspace*{-0.8cm}
\begin{center}
\begin{tabular}{@{\hspace{-0.25cm}}c}
\epsfxsize11.5cm\epsffile{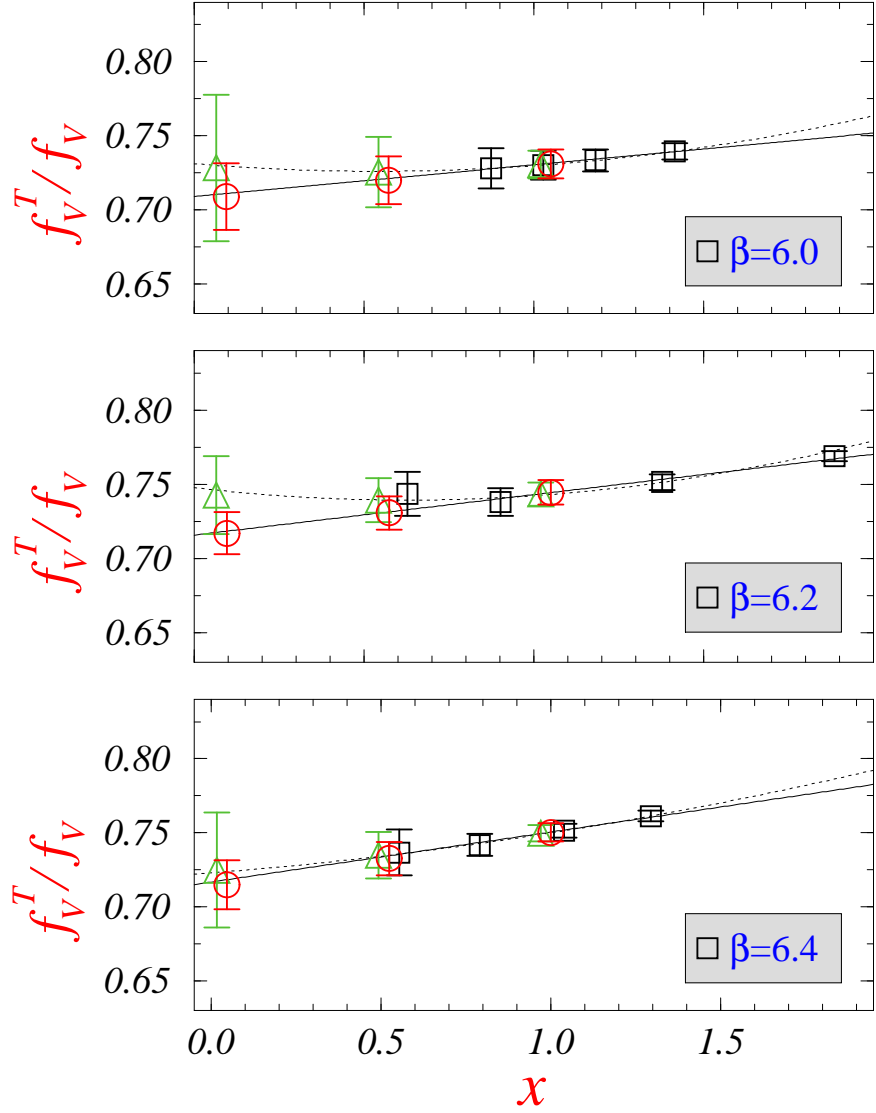}    \\
\end{tabular}
\vspace*{-.1cm}
%%%%%%%%%%%%%%%%%%%%%%%%%%%%%%%%%%%%%%%%%%%%%%%%%%%%%%%%%%%%%%%%%%
\caption{\label{fig:3}{\footnotesize 
Ratios $f^T_V(2\ \gev)/f_V$ that are directly accessed in our lattice study (denoted by squares) are 
fit linearly (solid line) and quadratically (dotted line) according to eq.~\eqref{extr}. To make 
the abscissa dimensionless we defined $x= (a m_P)^2/(a m_{ss})^2 $, where  
$(a m_{ss})^2=2(a m_K)^2-(a m_\pi)^2$. Resulting ratios for the $\rho$, $K^\ast$ and $\phi$ mesons 
are marked by circles. The ratios obtained from the quadratic fit (gray circles) are slightly 
shifted to the left as to make them discernible from the ones obtained 
through the linear fit (red circles).From left to right they correspond to $\rho$, $K^\ast$ and $\phi$ mesons.
 } }
%%%%%%%%%%%%%%%%%%%%%%%%%%%%%%%%%%%%%%%%%%%%%%%%%%%%%%%%%%%%%%%%%%
\end{center}
\end{figure}
Our final results are 
obtained through a linear extrapolation to the continuum limit ($a\to 0$), i.e.
\bea \label{extrapolate}
\left({f_V^T(2 \ \gev) \over f_V}\right) (a) = 
\left({f_V^T(2 \ \gev) \over f_V}\right)(0) + {\cal B} \ (a/r_0)^2 \;,   
\eea
where on the l.h.s. we use our results as obtained at three values of the lattice spacing, the values of 
$(a/r_0)$~\cite{necco} on the r.h.s. are given in table~\ref{tab:1}, and ${\cal B}$ is a fit parameter. 
Notice that since our calculation is improved at ${\cal O}(a)$, we omit the term linear in $a$ when  
extrapolating to the continuum limit.  The continuum extrapolation is illustrated 
in fig.~\ref{fig:4} and our final results are
\bea \label{results-final}
{f_\rho^T(2 \ \gev)/ f_\rho} &=& 0.720(24)\left(^{+16}_{-0}\right)\;, \nonumber \\ 
&& \cr
{f_{K^\ast}^T(2 \ \gev)/f_{K^\ast}}& =& 0.739(17)\left(^{+3}_{-0}\right)\;,\\   
&& \cr
{f_{\phi}^T(2 \ \gev)/f_{\phi}} &=&  0.759(9)(0) \;.\nonumber 
\eea
\par
\begin{table}[h!!] 
\begin{center} 
\begin{tabular}{|c|c|c|c|}
\hline
{\phantom{\Huge{l}}}\raisebox{-.1cm}{\phantom{\Huge{j}}}
$(a/r_0)^2$ & $f_\rho^T(2 \ \gev)/f_\rho$ & $f_{K^\ast}^T(2 \ \gev)/f_{K^\ast}$ & 
$f_{\phi}^T(2 \ \gev)/f_{\phi}$ \\ \hline \hline
{\phantom{\Huge{l}}}\raisebox{-.1cm}{\phantom{\Huge{j}}}
$0.035$  &  $0.708(23)\left(^{+19}_{-00}\right)$& $ 0.720(16)\left(^{+6}_{-0}\right)$   & $ 0.731(10)(0)$   \\
{\phantom{\Huge{l}}}\raisebox{-.1cm}{\phantom{\Huge{j}}}
$0.018$  &  $0.717(14)\left(^{+25}_{-0}\right)$    & $ 0.731(11)\left(^{+8}_{-0}\right)$   & $ 0.744(8)(0)$   \\
{\phantom{\Huge{l}}}\raisebox{-.1cm}{\phantom{\Huge{j}}}
$0.011$  &  $0.715(17)\left(^{+10}_{-0}\right)$    & $ 0.733(11)\left(^{+2}_{-0}\right)$   & $ 0.750(6)(0)$   \\ \hline
\end{tabular}
\vspace*{.8cm}
\caption{\label{tab:3}
\footnotesize  The values of the ratio of the vacuum-to-meson couplings mediated by the tensor 
vs. vector current, renormalized at $\mu = 2\ \gev$ in the (Landau)RI/MOM scheme, which 
at NLO in perturbation theory is the same as the $\msbar$(NDR) renormalization scheme.  }
\end{center}
\end{table}
\begin{figure}
\vspace*{-0.8cm}
\begin{center}
\begin{tabular}{@{\hspace{-0.15cm}}c}
\epsfxsize12.0cm\epsffile{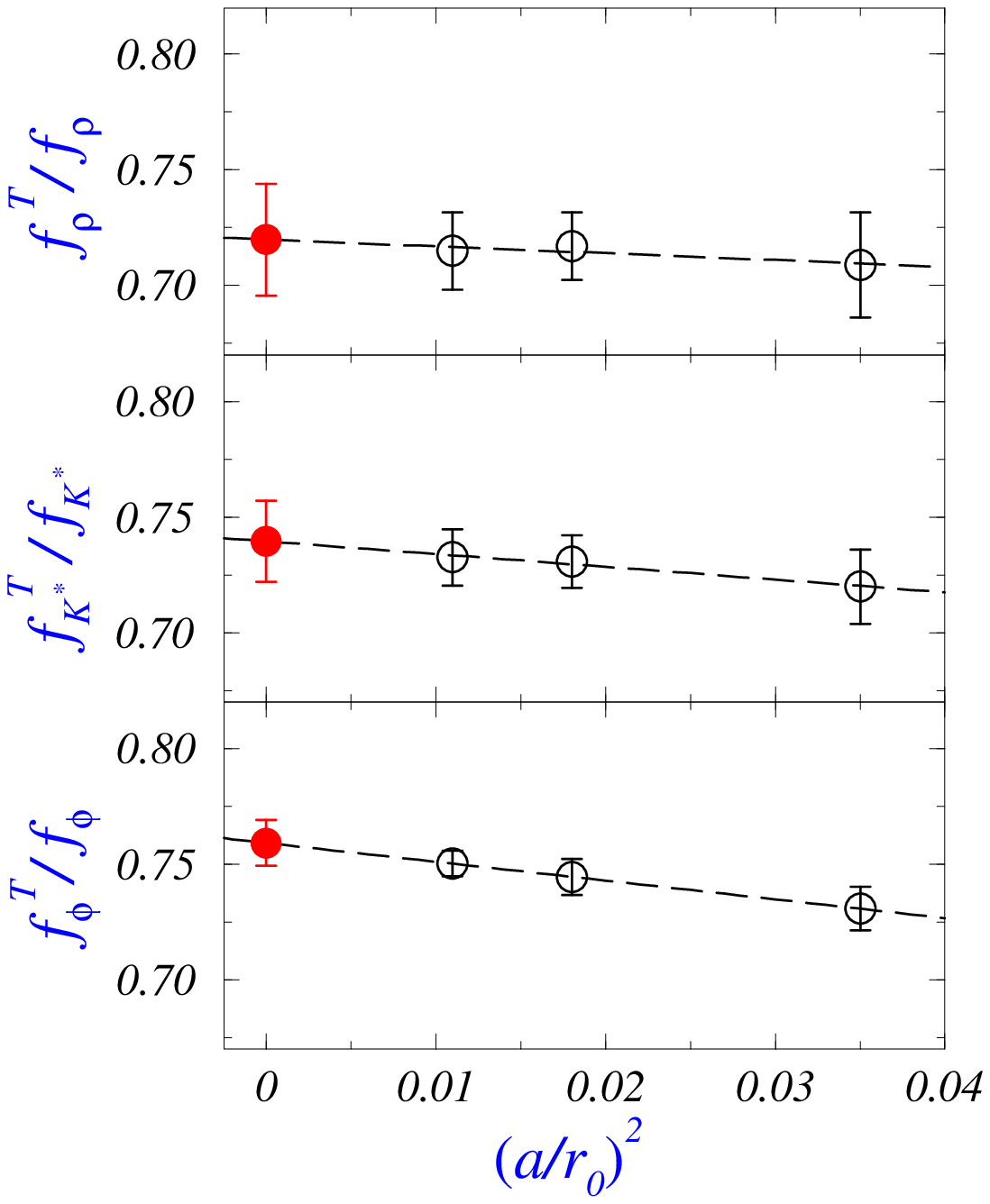}    \\
\end{tabular}
%%%%%%%%%%%%%%%%%%%%%%%%%%%%%%%%%%%%%%%%%%%%%%%%%%%%%%%%%%%%%%%%%%
\caption{\label{fig:4}{\footnotesize 
Extrapolation to the continuum limit ($a\to 0$).  
 } }
%%%%%%%%%%%%%%%%%%%%%%%%%%%%%%%%%%%%%%%%%%%%%%%%%%%%%%%%%%%%%%%%%%
\end{center}
\end{figure}

\subsection{A brief discussion of the systematic uncertainties}

Many systematic uncertainties cancel in the ratio of the vacuum-to-meson couplings 
computed here. 
For example, the uncertainty due to the lattice spacing is only present in the evolution 
of the tensor current from the renormalization scale $\mu=1/a$, down to $\mu=2$~GeV. 
As we showed, that uncertainty is at the level of a few per mil, thus  
negligible. 

Our central results at each value of $\beta$, presented in table~\ref{tab:3}, 
are obtained by fitting our data to the form~\eqref{extr} linearly ($\alpha_2 =0$). 
The second errors reflect the difference between those and the values 
obtained through the quadratic fit~\eqref{extr}. 
These errors are quite significant for the $\rho$-meson because that meson is reached 
through an extrapolation. On the other hand, they are small for the $K^\ast$- and $\phi$-mesons.

The systematic errors induced by the presence of the chiral 
logarithms in extrapolating to the chiral limit are unlikely to affect 
our ratios because the same hadron state appears in both the numerator 
and the denominator. One may also speculate that for the same reason 
 quenching effects should be small, but we prefer to consider our results as {\it quenched} 
and leave the unquenching issue to  future lattice studies.

In the continuum extrapolation we use a linear fit in the square of the lattice spacing since our 
results do not suffer from the ${\cal O}(a)$ artefacts. In fact all the renormalisation and improvement 
constants are known non-perturbatively, except for $b_T$ which is obtained from $1$-loop boosted 
perturbation theory. Since the residual ${\cal O}(a)$-effect is $\propto a m_q \alpha_s^2$ 
and we work with the light quarks ($(am_q)\leq 0.07$~\cite{qmass}), this can be safely ignored.

\section{Tensor coupling and the comparison with other theoretical predictions}

To get the absolute values of the tensor couplings we will now use our ratios from eq.~\eqref{results-final}, 
and multiply them by the physical values for $f_{\rho,K^\ast,\phi}$ extracted from experiments, i.e.
\bea \label{phenom}
f_{\rho,K^\ast,\phi}^T(2\ \gev) = \left({f_{\rho,K^\ast,\phi}^T(2\ \gev)\over f_{\rho,K^\ast,\phi}}\right)^{\rm latt.}
f_{\rho,K^\ast,\phi}^{\rm exp.}
\eea

\subsection{Our results for $f^{T}_{\rho,K^\ast,\phi}(2\ \gev)$ 
and comparison with other lattice calculations}

The experimental values for the charged vector meson couplings $f_{\rho^\pm},f_{K^{\ast \pm}}$ 
can be extracted from the data for $\tau$-lepton decays~\cite{PDG}:
\bea
&&BR( \tau \to \rho^- \nu_\tau) = 25.0(3)\%\;,\qquad BR( \tau \to K^{\ast -} \nu_\tau) = 1.29(5)\%\;.
\eea
By using the tree level expression
\bea \label{debole}
BR( \tau \to V^- \nu_\tau) = {G_F^2 m_\tau \vert V_{qq^\prime} \vert^2 \over 8 \pi} \tau_\tau m_V^2 f_V^2
\left( 1 - {m_\tau^2\over 2 m_V^2} \right) 
\left( 1 + {m_V^2\over m_\tau^2} \right)^2 \;,
\eea
and $\vert V_{ud}\vert =0.9735$, 
$\vert V_{us}\vert =0.220$, $\tau_\tau= 290.6$~ps, $m_\tau = 1.777$~GeV,  
$m_{K^{\ast }}= 0.892$~GeV, and $G_F=1.1664 \cdot 10^{-5}\ \gev^{-2}$, all taken from ref.~\cite{PDG}, 
we obtain
\bea \label{charged}
f_{\rho^\pm}^{\rm exp.} \simeq 208\ \mev\;,\qquad f_{K^{\ast \pm}}^{\rm exp.} \simeq 217~\mev\;,
\eea
where we display only the central values. 
Notice that the value of $f_{\rho^\pm}$ is consistent with $f_{\rho^0} = 216(5)$~MeV, obtained 
from $e^+e^-$-annihilation.

The constant $f_\phi$ can be obtained from the width $\Gamma_{\phi^0\to e^+e^-} = 1.32(4)$~keV, which 
we combine with $\alpha_{\rm em.}^{-1}=137.036$, $m_\phi=1.019$~GeV~\cite{PDG}
and get~\footnote{In the previous version of this paper, eqs.~\eqref{debole} and
\eqref{phi-em} contained typos. Numerical results however were correct.  
We thank Patricia Ball for drawing our attention to this. } 
\bea \label{phi-em}
f_\phi^{\rm exp.} = \left({ 27 m_\phi \over 4 \pi \alpha_{\rm em.}^2} 
\Gamma_{\phi^0\to e^+e^-}\right)^{1/2} \simeq 233~\mev\;.
\eea

We then insert the above values for $f^{exp}_{\rho,K^\ast,\phi}$ 
in~\eqref{phenom} and arrive at our estimate for the tensor couplings:
\bea  \label{tens-lattice}
&&f_{\rho^\pm}^T(2\ \gev)\  =\  150(5)\left(^{+3}_{-0}\right)~\mev\  =\  152(7)\  \mev\;, \nonumber \\
&& \cr
&&f_{K^{\ast \pm}}^T(2\ \gev)\  =\  160(4)\left(^{+1}_{-0}\right)~\mev\  =\  161(4)\ \mev\;,   \\
&& \cr
&&f_{\phi}^T(2\ \gev)\  =\  177(2)(0)~\mev\  =\  177(2)\ \mev\;. \nonumber  
\eea

The result for $f_\rho^T(2\ \gev)$ agrees also very well with the estimate of ref.~\cite{QCDSF} 
in which this constant has been computed on the lattice, at three values of the lattice spacing, 
in the quenched approximation. In that study  the perturbatively evaluated renormalisation constant has been used.
Their new calculation~\cite{QCDSF2} includes also the non-perturbative determination of $Z_T(\mu)$, 
and after extrapolating to the continuum limit they quote 
\bea
f_\rho^T(2\ \gev) = 150(4)\ \mev\,.
\eea
Although the systematic uncertainty of that result is still preliminary, we note a very 
pleasant agreement with our value given in eq.~\eqref{tens-lattice}~\footnote{We are indebted to 
Gerrit Schierholz for pointing out the ref.~\cite{QCDSF} to us and for communicating 
their new results.}.

\subsection{Comparison with the QCD sum rule results}

The QCD sum rule (QSR) estimates~\cite{bb2,bb} are made at the renormalization scale $\mu=1~\gev$. 
We run those values to $\mu = 2~\gev$, 
by using eq.~\eqref{evol} and $\Lambda_\msbar^{n_F=3} = 338(40)$~MeV~\cite{bethke}, i.e.
$f_V^T(2~\gev) = 0.9667(6)\times f_V^T(1~\gev)$. Thus the QSR estimates are
\bea \label{tens-qsr}
&&f_\rho^T(2~\gev)=155 \pm 10~\mev\;, \nonumber \\
&& \cr
&&f_{K^\ast}^T(2~\gev)=179 \pm 10~\mev\;,   \\
&& \cr
&&f_{\phi}^T(2~\gev)=208\pm 15~\mev\,. \nonumber 
\eea
Note also that the coupling $f_\rho^T(\mu)$ has been re-examined in ref.~\cite{bm}, 
essentially confirming the results of ref.~\cite{bb}. 
By comparing the lattice results~\eqref{tens-lattice} with the QSR values~\eqref{tens-qsr}, 
we see that the agreement of the two methods for $f_\rho^T$ is very good, quite 
good for $f_{K^\ast}^T$ and less good for $f_{\phi}^T$. Investigation of the source for 
that discrepancy is beyond the scope of the present paper. It should, however, 
be stressed that the LCSR yield results for full QCD while our results are quenched
and therefore a perfect match is anyway not expected.

\section{Summary}

In this letter we presented quenched lattice results for the ratio of the couplings 
of the light vector mesons to the tensor and to the vector currents. Those ratios 
enter the LCSR analyses of the phenomenologically important $heavy \to light$ 
meson semileptonic decay form factors.  From the results obtained with ${\cal O}(a)$ 
improved Wilson quarks, and with high statistics data, we were able to extrapolate 
to the continuum limit. The resulting ratios are then multiplied by the experimentally 
measured vector meson couplings to arrive at the tensor couplings. Comparison 
with the prediction obtained from the QCD sum rule analysis indicates good agreement 
for   $f_{\rho,K^\ast}^T(2\ \gev)$, whereas the agreement is not good in the case 
of $f_{\phi}^T(2\ \gev)$. Our result for $f_{\rho,K^\ast}^T(2\ \gev)$ also agrees very 
well with the lattice estimate of refs.~\cite{QCDSF,QCDSF2}.

\section*{Acknowledgements}
We thank G.~Martinelli and J.~Flynn for reading the manuscript, 
as well as Patricia Ball and G.~Schierholz for 
correspondance. 
This work has been partially supported by M.U.R.S.T, and by 
the E.C.'s contract HPRN-CT-2000-00145 ``Hadron Phenomenology from Lattice QCD".

\end{document}